\documentclass[usenatbib]{mn2e}
\usepackage{epsfig}
\usepackage{amssymb}
\usepackage{lscape,graphicx}
\usepackage{longtable}
\usepackage{lscape}
\usepackage{rotating}
\usepackage{times}
\usepackage{graphicx}
\usepackage{epstopdf}

\def\gsim{\mathrel{\raise0.35ex\hbox{$\scriptstyle >$}\kern-0.6em
\lower0.40ex\hbox{{$\scriptstyle \sim$}}}}
\def\lsim{\mathrel{\raise0.35ex\hbox{$\scriptstyle <$}\kern-0.6em
\lower0.40ex\hbox{{$\scriptstyle \sim$}}}}

\begin{document}
\vspace{-10cm}
\title[Far-Infrared excess around clusters in H-ATLAS]
{\textit{Herschel}-ATLAS: detection of a far-infrared population around galaxy clusters\thanks{{\it Herschel} is an ESA space observatory with science instruments provided by European-led Principal Investigator consortia and with important participation from NASA.}}
\vspace{-1cm}
\author[Coppin et al.]{ 
\parbox[t]{\textwidth}{
K.\ E.\ K.\ Coppin,$^{\! 1,2}$ J.\ E.\ Geach,$^{\! 1,2}$ Ian
Smail,$^{\! 2}$ L.\ Dunne,$^{\! 3}$ A.\ C.\ Edge,$^{\! 2}$ R.\ J.\
Ivison,$^{\! 4,5}$ S.\ Maddox,$^{\! 3}$ 
R.\ Auld,$^{\! 6}$ M.\ Baes,$^{\! 7}$ S.\ Buttiglione,$^{\! 8}$ A.\ Cava,$^{\! 9}$ D.L.\
Clements,$^{\! 10}$ A.\ Cooray,$^{\! 11}$
A.\ Dariush,$^{\! 6,12}$ G.\ De Zotti,$^{\! 8,13}$ S.\ Dye,$^{\! 6}$ S.\ Eales,$^{\! 6}$
J.\ Fritz,$^{\! 7}$ R.\ Hopwood,$^{\! 10}$ E.\ Ibar,$^{\! 4}$ M.\
Jarvis$^{\! 14,15}$ M.\ J.\ Micha{\l}owski,$^{\! 5}$ D.\
Murphy,$^{\! 2}$ M.\ Negrello,$^{\! 16}$ E.\ Pascale,$^{\! 6}$  M.\ Pohlen,$^{\! 6}$ E.\
Rigby,$^{\! 3}$ G.\ Rodighiero,$^{\! 17}$ D.\ Scott,$^{\! 18}$
S.\ Serjeant$^{\! 16}$ D.\ J.\ B.\ Smith,$^{\! 14}$ P.\ Temi,$^{\! 19}$ P.\ van der
Werf $^{\! 5,20}$}\\\\
$^{1}$ Department of Physics, McGill University, Ernest Rutherford
Building, 3600 Rue University, Montr\'{e}al, Qu\'{e}bec, H3A 2T8, Canada\\
$^{2}$ Institute for Computational Cosmology, Durham University, South
Road, Durham, DH1 3LE, UK\\
$^{3}$ School of Physics and Astronomy, University of Nottingham,
 University Park, Nottingham NG7 2RD, UK\\
$^{4}$ UK Astronomy Technology Centre, Royal Observatory, Blackford Hill, Edinburgh, EH9 3HJ, UK \\
$^{5}$ SUPA, Institute for Astronomy, University of Edinburgh, Royal
Observatory, Blackford Hill, Edinburgh, EH9 3HJ, UK\\
$^{6}$ School of Physics \& Astronomy, Cardiff University, Queens
Buildings, The Parade, Cardiff, CF24 3AA, UK\\
$^{7}$ Sterrenkundig Observatorium, Universiteit Gent,
 Krijgslaan 281 S9, B-9000 Gent, Belgium\\
$^{8}$ INAF -- Osservatorio Astronomico di Padova,
 Vicolo dell'Osservatorio 5, I-35122 Padova, Italy\\
$^{9}$ Departamento de Astrof\'{\i}sica, Facultad de CC. F\'{\i}sicas,
Universidad Complutense de Madrid, E-28040 Madrid, Spain\\
$^{10}$ Astrophysics Group, Physics Department, Blackett Lab, Imperial
College, Prince Consort Road, London SW7 2AZ, UK \\
$^{11}$ Department of Physics and Astronomy, University of California,
Irvine, CA 92697, USA \\
$^{12}$ School of Astronomy, Institute for Research in Fundamental
Sciences (IPM), PO Box 19395-5746, Tehran, Iran \\
$^{13}$ SISSA, Via Bonomea 265, I-34136 Trieste, Italy\\
$^{14}$ Centre for Astrophysics Research, Science \& Technology
Research Institute, University of Hertfordshire, Hatfield, Herts, AL10
9AB, UK \\
$^{15}$ Physics Department, University of the Western Cape, Cape Town,
7535, South Africa \\
$^{16}$ Department of Physics \& Astronomy, The Open University, 
Milton Keynes, MK7 6AA, UK\\
$^{17}$ Dipartimento di Astronomia, Universita di Padova,
 Vicolo Osservatorio 2, I-35122 Padova, Italy\\
$^{18}$ Department of Physics and Astronomy, University of British
Columbia, 6224 Agricultural Road, Vancouver, BC V6T 1Z1, Canada\\
$^{19}$ Astrophysics Branch, NASA Ames Research Center,
 MS 245-6, Moffett Field, CA 94035, USA\\
$^{20}$ Leiden Observatory, Leiden University, PO Box 9513, NL - 2300 RA Leiden, The Netherlands
}
\maketitle
\vspace{-6cm}

\begin{abstract}
We report the detection of a significant excess in the surface density of
far-infrared sources from the {\it Herschel}-Astrophysical Terahertz Large
Area Survey (H-ATLAS) within $\sim1$\,Mpc of the centres of 66
optically-selected clusters of galaxies in the SDSS with $\left<z\right>\sim
0.25$. From the analysis of the multiwavelength properties of their
counterparts we conclude that the far-infrared emission is associated with
dust-obscured star formation and/or active galactic nuclei within galaxies in
the clusters themselves. The excess reaches a maximum at a radius of
$\sim0.8$\,Mpc, where we find $1.0\pm0.3$ S$_{250}>$34\,mJy sources on average per cluster above
what would be expected for random field locations. If the far-infrared
emission is dominated by star formation (as opposed to AGN) then this
corresponds to an average star formation rate of $\sim$7\,M$_\odot$\,yr$^{-1}$
per cluster in sources with
$L_\mathrm{IR}>5\times10^{10}$\,L$_{\odot}$. 
Although lensed sources make a negligible contribution to the excess
signal, a fraction of the sources around the clusters could be
gravitationally lensed, and we have identified a sample of potential
cases of cluster-lensed \textit{Herschel} sources that
could be targeted in follow-up studies.\end{abstract}

\vspace{-1cm}
\section{Introduction}

The intimate connection between galaxies' environments and their
star formation histories is evident in the evolution of the cluster galaxy
stellar mass function (e.g.~\citealt{Vulcani10}). The most massive galaxies in
rich clusters today appear to have been in place in progenitor environments
since at least $z\sim1$ (\citealt{dePropris99}; \citealt{Kodama03};
\citealt{Neistein06}), and there has been little evolution in the
number density of the most massive elliptical galaxies in the intervening 8\,Gyr
\citep{Balogh01}. However,
significant stellar mass evolution is still required in the remainder of the
cluster population during this period \citep{Balogh01}. The key features of
this evolution are:  (a) a steepening of the faint-end towards $z=0$, that is, low-mass end of
the luminosity function \citep{Stott07}; and (b) the appearance of a population of
passive, massive lenticular (S0) galaxies in the cores of clusters since
$z\sim0.5$ \citep{Dressler97}.

The increase in the fraction of low-mass galaxies can be explained by the
continuous accretion of satellite galaxies (with subsequent gas stripping and
the cessation of further cooling preventing further growth). The formation of
S0s can only be accounted for in an evolutionary sequence connecting distant
gas-rich discs undergoing a period of additional star formation to
build up the total stellar mass and enhancement of bulge-to-disc ratios
(\citealt{Poggianti99}; \citealt{Kodama01}).  Until relatively recently,
evidence for the large star formation rates (SFRs) required for such a
transformation in the spiral populations of intermediate redshift clusters was
lacking. However, since the advent of sensitive mid- and (now) far-infrared
panoramic surveys, several studies have now revealed a population of hitherto
optically hidden star-forming galaxies in rich clusters over $0.3\lesssim
z\lesssim 1.5$ (e.g.~\citealt{Geach06}; \citealt{Duc00,Duc04};
\citealt{Fadda00}; \citealt{Metcalfe03}; \citealt{Finn10};
\citealt{Kocevski10}; \citealt{Braglia10}).

It has become clear that these obscured star-forming galaxies could be
responsible for strong evolution in the stellar mass function of even rich
clusters since $z\sim0.5$. Thus, not only does this population provide a key
insight into various environmental effects on the star formation histories of
relatively `normal' galaxies, but it also represents an important stage in the
overall shaping of the galaxy population today. The globally-averaged
total SFR in rich clusters as well as the average field has been in strong decline since
$z\sim0.5$, although it is unclear how the strength of the evolution
is tied to galaxies' environments. Indeed infrared studies have revealed significant variation in the SFRs of individual
clusters \citep{Geach06}. It is thought that the origin of this variation
could be rooted in the different environmental conditions specific to
different clusters, such as sub-structure, dynamical state, thermodynamic
properties of the intracluster medium (ICM), etc. The next step in
understanding this variation, and building up a more statistical picture of
the evolution of the obscured star-forming populations of clusters since
$z\sim0.5$, is to turn to wide-field panoramic infrared surveys of a much
larger sample of clusters and groups covering a large dynamic range of
environment.

While previous surveys undertaken with \textit{Spitzer} and the
\textit{Infrared Space Observatory (ISO)} have mapped the mid-infrared
emission (e.g.~15--24\,$\mu$m) of clusters, panoramic far-infrared surveys
have so far been challenging. Both ground- and space-based surveys have lacked
the field-of-view, sensitivity and resolution to cover large areas down to the
required depths to pin-point the obscured star-forming galaxy population
(e.g.\ \citealt{Wardlow10}). The {\it Herschel} space telescope
\citep{Pilbratt10} has enabled us to move beyond these limitations
(e.g.~\citealt{Rawle10}).

The \textit{Herschel}-Astrophysical Terahertz Large Area Survey (H-ATLAS;
\citealt{Eales10}) is the widest-area submillimetre
\textit{Herschel}-SPIRE \citep{Griffin10} and PACS \citep{Poglitsch10}
survey, and -- when complete -- will cover an area of
$\sim550$\,deg$^{2}$ from 100--500\,$\mu$m. The large volume probed will
contain a large number of galaxy clusters, and the sensitivity of the far-infrared
observations will allow us to systematically search for obscured star-forming
galaxies in their vicinity. This letter presents a statistical
analysis of the \textit{Herschel} SPIRE sources in
the core of $0.07<z<0.43$ clusters as mapped by the Science Demonstration
Phase (SDP) H-ATLAS observations, covering a $\sim14.4$\,deg$^{2}$ field at
9$^{\rm h}$ (\citealt{Pascale10}; \citealt{Ibar10}). 
Our goals are twofold: (a) to search for \textit{statistical}
evidence of dust-obscured star formation activity in this cluster
population; and (b) to identify any candidate cluster-lensed sources for further
study and follow-up.

This paper is organised as follows:  we describe our
unique cluster detection algorithm and the H-ATLAS SDP SPIRE catalogue in
\S~\ref{sample}, the statistical analysis and results of the H-ATLAS
and cluster catalogue cross-correlation in \S~\ref{analysis}, and
summarise our findings in \S~\ref{summary}. Throughout we assume
cosmological parameters from the \textit{WMAP} fits in
\citet{Spergel03}: $\Omega_\Lambda=0.73$, $\Omega_\mathrm{m}=0.27$,
and $H_\mathrm{0}=71$\,km\,s$^{-1}$\,Mpc$^{-1}$.

\section{Cluster and H-ATLAS catalogues}\label{sample}
\subsection{Cluster detection}\label{cluster}

\begin{figure}
\epsfig{file=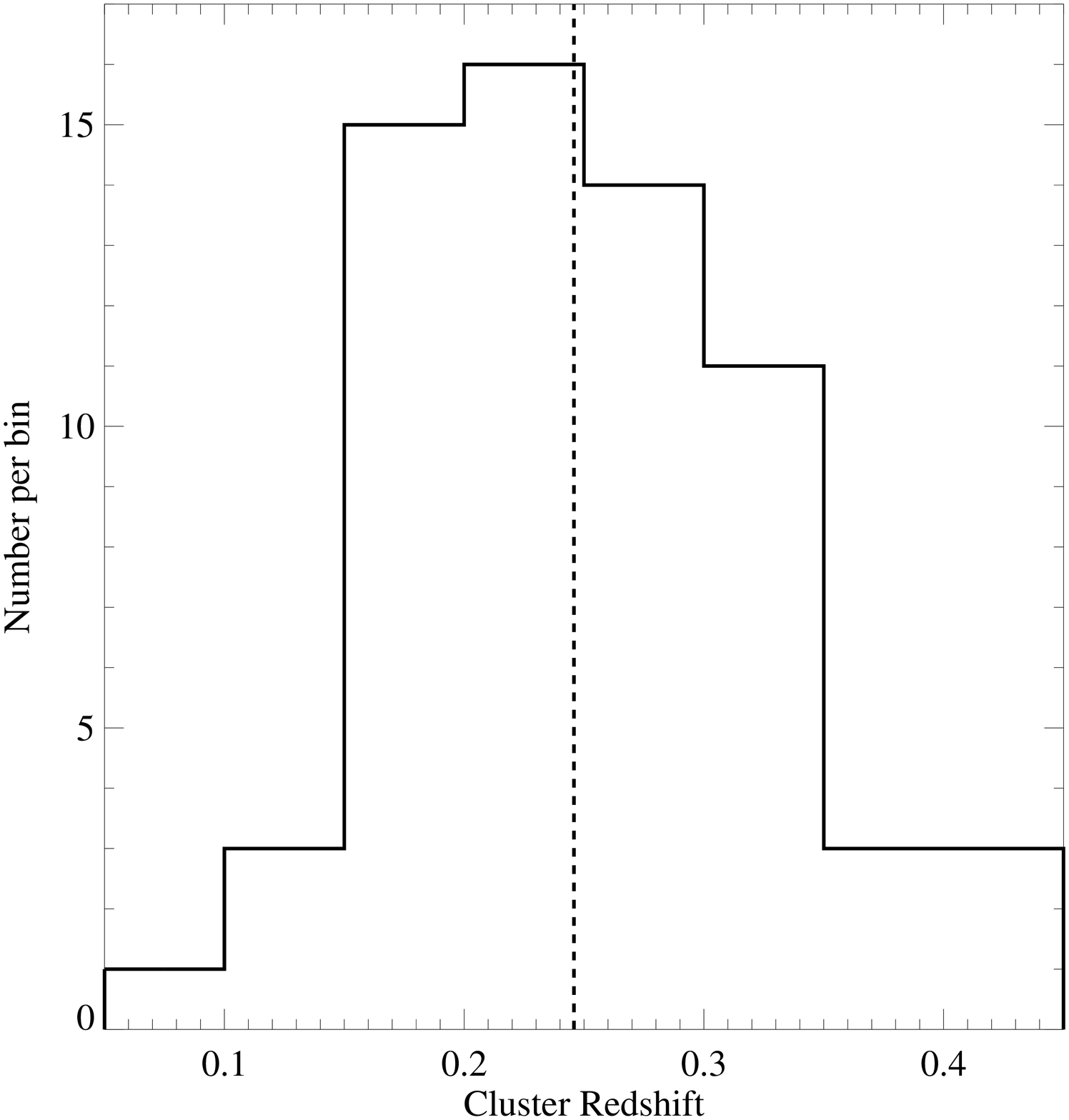,width=0.5\textwidth}
\caption{The redshift distribution of our 66-member cluster sample
  within the H-ATLAS SDP coverage of $\simeq14.4$\,deg$^{2}$. We have
  estimated the cluster redshifts using the photometric (and for 6.5\%
  of the time spectroscopic) redshifts of the 549 cluster
member galaxies (\citealt{Abazajian09}). The clusters span a redshift of 0.07--0.43
85\% of the sample are at $0.15\leq z\leq 0.35$), and have a median redshift
of $\left<z\right>=0.25$ (which we have indicated with a vertical dotted line) at which the angular scale is 240\,kpc
arcmin$^{-1}$.} \label{fig:histogram} \end{figure}

We have used the technique presented in \citet{MGB10} to identify clusters of
galaxies from panoramic optical imaging. Briefly, the method uses a series of colour
selections to first isolate `red-sequence' cluster members (those where the
4000\AA\ break is bracketed by two filters), followed by the construction
of a Voronoi diagram of the projected galaxy distribution. Clusters and
groups are identified as associations of Voronoi cells, sharing at least one
vertex between cells, with areas significantly lower (i.e.\ higher galaxy
surface densities) than would be expected if the galaxies were randomly
distributed on the sky. In this case we used photometry from the Sloan Digital
Sky Survey \citep{York00} 7th Data Release (SDSS DR7; \citealt{Abazajian09}).
For the selection, we employed Galactic extinction corrected {\tt modelMag} in
the $(g-r)$, $(r-i)$ and $(i-z)$ bands; see \citet{Gunn98} for a description
of the SDSS photometric system. The minimum number of `connected' galaxies
that qualify as a cluster is five.  The position of the cluster core is
defined as the average of the positions of the member galaxies'
Voronoi cells, however we also define a `brightest cluster galaxy (BCG) centre' as the
location of the brightest cluster member in the $r$-band. Apertures placed on
the cluster core defined by the geometric mean provide the most
complete coverage of the member galaxies.  We find 66 clusters within
the H-ATLAS SDP coverage of $\simeq14.4$\,deg$^{2}$. 
The redshifts of the clusters have been estimated from
the photometric (and in some cases spectroscopic) redshifts of the cluster
members (\citealt{Abazajian09}).  We note that 36 members out of a total of 549
galaxies across all clusters have spectroscopic redshifts from SDSS
(6.5\%).  For sources with spectroscopic redshifts, the mean spectroscopic-to-photometric
  redshift offset is -0.0018, with a standard deviation of
  0.017.  Further details can be found in \citet{Geach11}. The clusters span a redshift of 0.07--0.43
85\% of the sample are at $0.15\leq z\leq 0.35$), and have a median redshift
of $\left<z\right>=0.25$ at which the angular scale is 240\,kpc
arcmin$^{-1}$ (see Fig.~\ref{fig:histogram}).
 Based on tests performed on mock
catalogues, the cluster catalogue is $>$90\% complete at a halo mass of
$10^{14}$\,M$_\odot$ \citep{MGB10}.  The number of false positives can be estimated by randomly shuffling the
colours of galaxies (while keeping the positions fixed) and re-running the
detection algorithm. At the lower membership limit, the number of false
detections is expected to be 0.06\,deg$^{-2}$ or $<1$ of the 66 clusters.
Further details of the cluster algorithm, selection and completeness can be
found in \citet{MGB10}.  We estimate the cluster
richness using the commonly used $B_{\rm gc}$ statistic, an
approximation of the amplitude of the real-space correlation function \citep{Longair79}.  
\citet{Yee03} show that this statistical measure is reasonably well
correlated with the physical properties of the clusters, and we apply these scalings
to find the typical cluster scale $R_{200}$\footnote{$R_{200}$ is the
    equivalent radius enclosing a density $\geq200\times$ the critical density.}$\simeq(1.2\pm0.4)$\,Mpc and
$\log M_{200}/$M$_{\odot}$\footnote{$M_{200}$ is the mass within
  $R_{200}$.}$\simeq(14.7\pm0.5)$, although the errors on individual $B_{\rm gc}$ measurements are large.

\subsection{The H-ATLAS SDP catalogue}\label{spirecat}

The H-ATLAS SDP catalogue consists of 6876 sources detected at $>5\,\sigma$ in
either of the 250, 350 or 500\,$\mu$m bands over a $\simeq14.4$\,deg$^{2}$
region \citep{Rigby10}. The 5-$\sigma$ point source sensitivity limits (including
confusion noise) are 34, 38, and 44\,mJy at 250, 350, and 500\,$\mu$m,
respectively.  \citet{Smith10} have employed a likelihood ratio (LR) method to perform the
optical cross-identifications of the 6621 250\,$\mu$m-detected sources
with the SDSS DR7 catalogue with a limiting \emph{r}-band magnitude of
22.4 \citep{Abazajian09}.  The LR technique assigns a reliability, $R$, to each match 
and indicates the probability that the counterpart is the correct
identification. Of the 6876 H-ATLAS sources, 2423 are thus classified as
having a reliable ($R\geq0.8$) optical counterpart, and the remaining 4453
 as optically unidentified ($R<0.8$ or no optical counterparts).

\section{Analysis and Results}\label{analysis}
\subsection{Measurement of far-infrared emission around the clusters}\label{sec1}

\begin{figure}
\epsfig{file=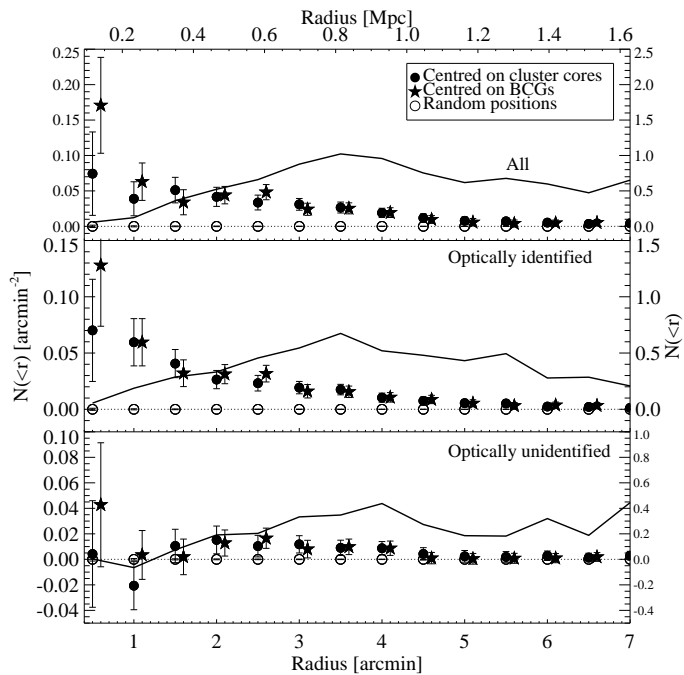,width=0.5\textwidth}
\caption{Cumulative number density of the average background-subtracted (points; axis on the
  LHS) and number (curve; axis on the RHS) of all (top panel),
  optically identified (middle panel), and optically unidentified
  (bottom panel) H-ATLAS sources as a function of clusto-centric radius. 
For comparison, we have also plotted the
average (background-subtracted) cumulative surface density as a function of radius from
the central BCG (stars).  The standard error of the mean has been used to
calculate the error bars. This plot shows that we have measured an
excess of H-ATLAS sources (with the significance varying with radius) 
within a projected clustocentric radius of 5$'$
(which corresponds to 1.2\,Mpc at $z=0.25$) over
that of the background field H-ATLAS sources, with optically identified sources (presumably
far-infrared-bright cluster members, see text) making the most significant
contribution to the measured excess. For reference, the average area of a
single cluster is 7.5\,arcmin$^{2}$, where the area is defined to contain 80\% of
the cluster members originally assigned by the Voronoi tessellation
detection.} \label{fig1} \end{figure}

The first step of our analysis is to simply measure the surface density of
H-ATLAS sources (both optically identified and unidentified) as a
function of projected clusto-centric radius around the 66
clusters (Fig.~\ref{fig1}). As a field control sample, we repeat this
exercise 1000 times for a set of 66 random positions across the field. As expected, at
large radii the surface density around the clusters is indistinguishable from
the average `field' estimate, however there is a clear positive excess of
far-infrared sources within $\sim5'$ (1.2\,Mpc for $z=0.25$) of the clusters, the
significance of which peaks at $\sim3.5'$. There is an average excess of
$\sim1$ source per cluster over the background, although
note that by definition the cluster environments are characterised by
an excess surface density of galaxies. The total number of H-ATLAS sources
detected within $3.5'$ of the 66 clusters is 401, representing a
$\simeq3.5\,\sigma$ excess of $67\pm20$ sources (the error is Poisson) above the background
signal of $332\pm1$ sources
on average (the error is the standard error of the mean).  At a radius of $5'$ from the 66 clusters, we find 719
sources (a less significant excess of $41\pm27$ sources over our Monte Carlo estimated
background signal of $678\pm1$ at the same clustocentric radius).  
For comparison, we have also repeated the above analysis using the projected radius from
the BCG as the cluster centre, and the signal in the $r<0.5'$ bin clearly
increases (see Fig.~\ref{fig1}) -- with six H-ATLAS sources
  lying within 8\,arcsec of BCGs (note that the 250\,$\mu$m PSF is
  19\,arcsec). This suggests that several H-ATLAS sources are
associated with the BCGs, either by lensing a background far-infrared source or
that the far-infrared emission is from the BCG itself,
e.g.~\citet{Edge10}. We have quantified the likelihood of finding this excess signal by
chance by using our Monte Carlo simulations and find that for 
radii $\lesssim3.5'$ (where the maximum excess signal occurs) we would
expect to see our average detected surface density $<0.1$\% of
the time in randomly sampled apertures of equivalent size in the field.
The simulations also reveal that at radii larger than about $5'$ the 
random chance of detecting our measured
surface density (or greater) near the clusters above the background becomes $>1$\%
and increases rapidly beyond $5'$.  Thus, for the following statistical
analyses we use the 719 H-ATLAS sources found within $5'$ of the 66
clusters, which strikes a good balance of identifying the majority of
the sources responsible for the excess signal while keeping the 
background field contribution to the signal to a minimum.

 We have calculated the surface density of H-ATLAS sources in
  angular bins, regardless of the individual cluster redshift. An
  alternative approach would be to calculate the surface density as a
  function of physical projected radius, which would be important for
  broad redshift distributions.  We conducted such an analysis as a check,
  by counting H-ATLAS sources within variable angular radii corresponding to a
  particular physical scale around the clusters.  We calculate the field
  estimate using the same Monte Carlo technique as above, but using 66
  apertures randomly drawn from a distribution function matching the
  cluster redshift distribution, repeating this 1000 times (see e.g.~\citealt{Temporin09}). The
  resulting surface density profile closely matches that found for
  angular bins, which is not surprising given our narrow redshift
  distribution (see Fig.~\ref{fig:histogram}), with the excess signal arising within $\simeq1$\,Mpc of the
  cluster cores, in agreement with the `average' physical scale shown
  in Fig.~\ref{fig1}.

There are two distinct physical origins for the excess signal: (a) obscured
star formation or active galactic nuclei (AGN) in cluster members; and (b) gravitational lensing of
background sources. The majority of the H-ATLAS sources in low
redshift clusters are expected to have
optical counterparts, whereas H-ATLAS sources with no robust counterpart are
most likely to be at higher redshift (except in the case of galaxy-galaxy
lensing, where the foreground lensing galaxy is identified as the
counterpart).  Splitting the sample into optically identified and
unidentified sub-samples therefore provides a crude method of determining if
the excess signal seen around clusters comes from the cluster members
themselves, or from strongly lensed background sources. From the 719
H-ATLAS sources found within 5$'$ of the 66 clusters, we find that 268 (37\%)
have optical counterparts with the remaining 451 (63\%)
having no optical counterpart. To examine the relative contribution to the far-infrared
excess signal, we repeat the radial surface density analysis described above
for these two sub-samples separately. Fig.~\ref{fig1} shows that the majority of
the excess signal seen in the full sample is due to the {\it optically identified} H-ATLAS
sources; while the surface density of optically unidentified H-ATLAS sources around clusters is
essentially statistically indistinguishable from the random field.   
We thus now focus our attention on the optically identified
sources around the clusters and defer a discussion of the nature of the
optically unidentified sources to Section~\ref{unid}.

\subsection{Is the far-infrared excess physically associated with the clusters?}
\label{excess}

We have determined that the excess signal of far-infrared emission around the
low-redshift cluster sample comes from H-ATLAS sources with robust
optical counterparts. These
could be cases where a foreground galaxy is lensing a background source
\citep{Negrello10} (where the lensing galaxy is the optical counterpart of the
H-ATLAS source), or the far-infrared emission is from the galaxy itself. Is there any evidence to suggest that the
majority of the optically identified H-ATLAS sources have redshifts
consistent with the clusters? 

We now perform a test to search for evidence that the far-infrared
colours of H-ATLAS sources around the clusters are consistent with the
redshift of the clusters. The optical photometric redshifts of the optical counterparts of the
H-ATLAS sources are not useful for reliably distinguishing galaxy-galaxy lensing from
the cluster members, since it is likely that in the case of galaxy-galaxy lensing, a far-infrared
source would be identified with the lensing galaxy instead of the true
background lensed source (which would be too faint/obscured to
be seen in the optical). Follow-up millimetre studies that can
positively identify the molecular gas emission of the high-redshift
source, unambiguously separating it in redshift space from the foreground galaxy, is
arguably the best technique. In lieu of those data, we can crudely use
the far-infrared colours as a rough redshift discriminator, since the
250, 350, and 500\,$\mu$m bands sample near the dusty spectral energy
distribution (SED) peak.  Thus,
following e.g.~\citet{Amblard10} we compare the
$S_\mathrm{250}/S_\mathrm{350}$ and
$S_\mathrm{500}/S_\mathrm{350}$ SPIRE
colours of the optically identified versus unidentified H-ATLAS
cluster-matched sources to
test if the optically identified sources are more consistent with
being lower redshift cluster members and the optically unidentified sources more
consistent with being background or cluster-lensed higher redshift
sources.  
We note that not of all the H-ATLAS sources have direct
detections in all three SPIRE bands (and thus errors on the
colours of those individual sources will be large).
Additionally, we note that $\sim27$\% of the 500\,$\mu$m sources are likely
blends of multiple sources (within a relatively large beamsize of
$\sim35''$), 
with their flux boosted by up to a factor of $\sim2$
\citep{Rigby10}.  We thus place more emphasis on the results involving the
250\,$\mu$m flux densities, which should suffer less from these effects due to
the relatively smaller beamsize ($\sim18''$), but use the 500\,$\mu$m
flux densities as a consistency check.  

We use a Kolmogorov-Smirnov (KS) test to compare the 
$S_\mathrm{250}/S_\mathrm{350}$ and
$S_\mathrm{500}/S_\mathrm{350}$ colours of the optically identified
subset of 268 cluster-matched H-ATLAS sources with the 451 optically
unidentified cluster-matched H-ATLAS sources to search for evidence of any differences between the colour distributions that would hint
at an overall redshift difference between the two samples. 
The KS test reveals that there is a $<1\times10^{-8}$ chance that they are drawn from the same
distribution. 
The optically identified cluster-matched H-ATLAS sources
have bluer $\left< S_\mathrm{250}/S_\mathrm{350}
\right> = 1.88\pm0.06$ colours and bluer $\left< S_\mathrm{500}/S_\mathrm{350} \right> = 0.4\pm0.02$ colours
on average than the optically unidentified cluster-matched H-ATLAS 
sources ($\left< S_\mathrm{250}/S_\mathrm{350}
\right> = 1.3\pm0.03$; 
$\left< S_\mathrm{500}/S_\mathrm{350}\right> =
0.5\pm0.01$).  
If we assume the median
$3\times10^{10}$\,L$_{\odot}<L_\mathrm{dust}<10^{11}$\,L$_{\odot}$
UV-submm template from Smith et al.\ (in preparation) for a `typical'
H-ATLAS galaxy SED, then these mean colours roughly indicate that the optically identified cluster-matched H-ATLAS
sources, on average, lie at redshifts of $z\sim0.25$ (consistent with the
cluster redshifts) and that the optically unidentified sources are at
typically higher redshifts of $z\sim1$.  

We thus conclude that the origin of the far-infrared excess signal around the
clusters originates from sources within the clusters themselves. These
galaxies are likely to be obscured star-forming galaxies or AGN. We note that
although the actual AGN content of clusters as a function of time is still
fairly poorly constrained, there is evidence that the infrared emission of the
general cluster population (i.e. galaxies on the outskirts of clusters) is
generated by star formation \citep{Geach09}.

We do expect some instances of galaxy-galaxy lensing in the optically
identified sample. Cases of galaxy-galaxy lensing could potentially be enhanced around clusters
simply due to the increased surface density of foreground galaxies, and the
increased mass density cross-section in cluster regions can boost
amplification further\footnote{For example, although rare, the Cosmic Eye
\citep{Smail07} is a $z\sim3$ galaxy lensed in a near-perfect Einstein ring by
a $z=0.7$ elliptical that meets the threshold for strong lensing \textit{only}
because we are viewing the $z=0.7$ galaxy through a foreground $z=0.3$ cluster
$\sim1'$ away.}. Therefore, a lensing galaxy may have a clear optical
identification as a background source, but could be lensed further due to the
presence of the cluster along the line of sight. We note that out of five
instances of strongly lensed optically identified H-ATLAS sources identified
by \citet{Negrello10}, one of these lies within $5'$ of a cluster core. This
is strictly a lower limit on the occurrences of galaxy-galaxy lensing in our
sample, however, the actual number of such cases is not expected to dominate
the optically identified SPIRE sources in our sample, given their typical
500$\mu$m fluxes and far-infrared colours described above.

\subsection{Integrated far-infrared emission from clusters at $z\sim0.25$}

\begin{figure}
\epsfig{file=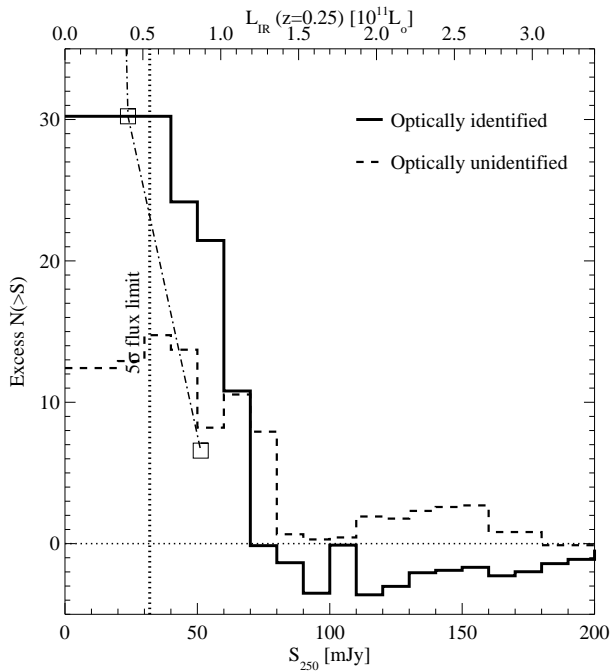,width=0.5\textwidth} 
\caption{Background-subtracted cumulative number counts of optically identified and unidentified H-ATLAS
sources within $5'$ of the 66 cluster cores.  
The 250\,$\mu$m 5\,$\sigma$ flux limit
of the H-ATLAS catalogue is shown by the dotted vertical
line. Based on the far-infrared colours, we expect
most of the signal from the optically identified sources here to be cluster
members (see text). As a guide, the upper scale corresponds to the total 8--1000\,$\mu$m
integrated luminosity, evaluated from the median
$3\times10^{10}$\,L$_{\odot}<\mathrm{L_{dust}}<10^{11}$\,L$_{\odot}$
UV-submm template from Smith et al.\ (in preparation) for a `typical'
H-ATLAS galaxy SED, redshifted to the
median redshift of the cluster sample ($z=0.25$), and normalised to the
250$\mu$m flux.  The $0.2<z<0.4$ \textit{Herschel} Multi-tiered
Extragalactic Survey (HerMES) submm luminosity function
(square symbols connected with a dot-dash curve; \citealt{Eales10Hermes}) is overplotted for comparison and has been scaled arbitrarily to the bin at our survey
flux limit.}\label{fig2} \end{figure}

If we now assume that the detected far-infrared emission is due to
obscured star formation (as opposed to AGN) within the
clusters then we can use the background-subtracted
luminosity function to estimate the average level of star formation in
the $z\sim0.25$ clusters. 
We integrate a background-subtracted histogram of the 250\,$\mu$m flux density
of the optically identified cluster-matched H-ATLAS sources (see
Fig.~\ref{fig2}) to yield a total
flux contribution of 1.6\,Jy over the 66 clusters, or $\sim$24\,mJy
per cluster. Assuming this is a proxy for dusty star formation we can convert this
to a total SFR by estimating the integrated total
(8--1000\,$\mu$m) infrared luminosity of the median H-ATLAS galaxy
template with
$3\times10^{10}$\,L$_{\odot}<L_\mathrm{dust}<10^{11}$\,L$_{\odot}$
from Smith et al.\ (in preparation) 
redshifted to $z=0.25$ (which is
well-matched to our sample far-infrared colours; see
Section~\ref{excess}) and 
normalised to the relevant 250\,$\mu$m flux density. We find an
average SFR per cluster of $\sim7\pm{3}$\,M$_\odot$\,yr$^{-1}$, applying the SFR calibration
of 8--1000\,$\mu$m integrated luminosity of \citet{Kennicutt98}. The
error range on the estimate represents the inferred difference in
total luminosity within the redshift range of $0.2<z<0.3$.  We note
that a systematic uncertainty in the SFR estimate comes from our
assumed SED, although this should not be a significant effect, since we
have used an SED template based on the H-ATLAS galaxies themselves
within this redshift range.  
Nevertheless, for example if an Arp\,220 or M82 SED is assumed (which have been shown to be inappropriate for our sample; see Smith
  et al.\ in preparation) we find an SFR systematically higher by a factor of $\sim6$ or $\sim10$, respectively.  The total SFR of the
clusters should also be considered a lower limit, given the 250\,$\mu$m sensitivity; since we only probe
down to $L_{\rm IR}\sim5\times10^{10}$\,L$_\odot$, we also expect a
contribution from dust-obscured galaxies below this limit. For
example, assuming the shape of the HerMES 250\,$\mu$m luminosity
function (\citealt{Eales10Hermes}; see Fig.~\ref{fig2}), if we could probe to a flux limit of
$\sim10$\,mJy we would expect an additional $\sim90$ sources down to
$\sim3\times10^{10}$\,L$_\odot$, contributing an \textit{additional}
$\sim7$\,M$_\odot$\,yr$^{-1}$ per cluster.

How does this compare to the total cluster-integrated SFR derived from
the UV alone?  To estimate the
UV integrated SFR for the clusters, we evaluate the total SDSS $u$-band flux density within 5$'$
of each cluster (corrected for Galactic extinction), and perform a field
correction based on the total flux within randomly placed apertures across the
field. We only sum the excess in galaxies not on the red-sequence (as these
are not present in the field), and derive an average total cluster
integrated SFR of $\sim$20\,M$_\odot$\,yr$^{-1}$, assuming the
\citet{Kennicutt98} SFR calibration. Thus, the total
  UV-derived rate appears to be comparable to the total far-infrared-derived rate, but of
  course with the caveats that (a) the UV-derived value is likely integrated further
  down the cluster luminosity function due to higher sensitivity to
  star formation in lower mass systems, and indeed includes many galaxies
  with low activity missed by the far-infrared survey, and (b) there is a large
  uncertainty in a simple conversion between the monochromatic
  $u$-band to SFR, and (c) we have not corrected for
  potentially non-negligible $u$-band emission from stars not associated
  with new star formation in red galaxies in the clusters (although we
  excluded the main red sequence).

With these caveats in mind, it is
  clear that the far-infrared-derived integrated SFR provides a superior
  estimate of the total level of star formation activity in the
  clusters, with the main uncertainty being the calibration of total
  far-infrared luminosity to SFR. However, it is also evident that the current
  limits of this H-ATLAS survey are missing part of the low-level
  activity in the clusters, as revealed by our simple UV estimates and
extrapolation of the HerMES luminosity function. Our measurements
should therefore be taken as lower limits to the integrated SFRs of
these clusters, as mentioned previously.

\begin{figure} 
	\epsfig{file=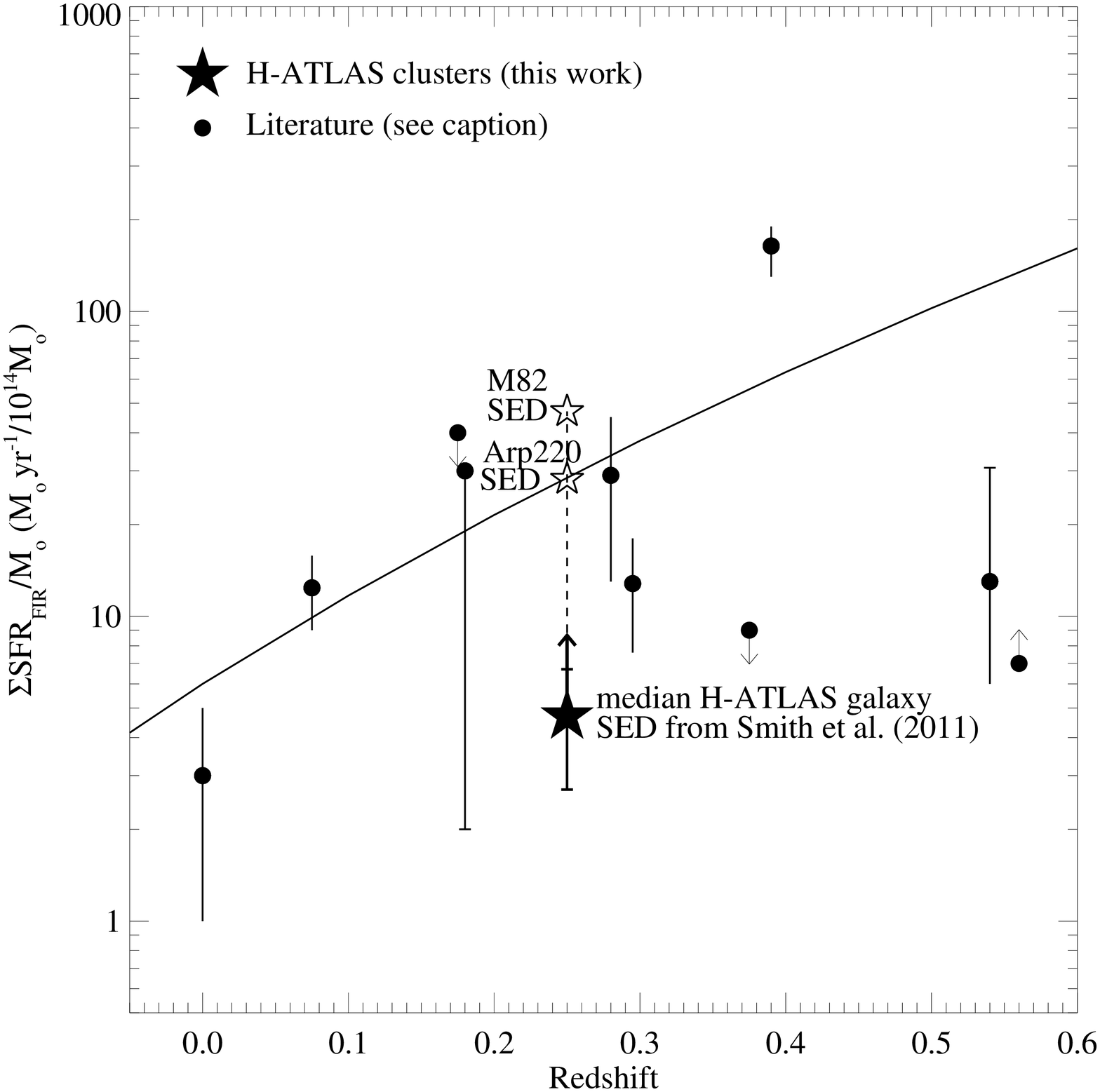,width=0.5\textwidth}
\caption{Measures of the mass-normalised SFRs in galaxy clusters out to
$z\sim0.6$, adapted from fig.~6 of \citet{Geach06}. The SFRs are derived from
the mid- or far-infrared populations within $\sim$2\,Mpc and are normalised to
the best estimate of the total (luminous+dark) cluster mass. Listed in order
of increasing redshift, they are: Perseus \citep{Meusinger00}, A3112
\citep{Braglia10}, A2218 \citep{Biviano04}, A1689 (\citealt{Fadda00};
\citealt{Duc02}), H-ATLAS (this work), A1758 \citep{Haines09}, the Bullet
cluster (\citealt{Rawle10}; \citealt{Clowe04}; \citealt{Markevitch06}), A370
\citep{Metcalfe05}, Cl 0024+16 and MS0451--03 \citep{Geach06}, and J1888.16LC
\citep{Duc04}. An evolutionary model for the counts of star-forming ULIRGs
from \citet{Cowie04} is overlaid as a guide only, and has been normalised
arbitrarily to the mean star formation rate in Cl\,0024+16 and MS\,0451$-$03
from \citet{Geach06}. This plot shows an increasing rate of activity in more
distant clusters as traced through their mid- or far-infrared populations,
albeit with a large scatter, suggestive that the infrared is a sensitive
tracer of environmental changes within the clusters. Note that there are large
systematic effects on the points in this plot, with the inferred SFR dependent
on the assumed form of the SED, and estimate of the total cluster mass. Given
the large uncertainties on the H-ATLAS cluster masses and the cool SED we have
adopted, we consider our point a lower limit (as indicated by the arrow). We
also show how our result changes if we instead assume an M82 or Apr220 SED
(open symbols), to facilitate a comparison with the literature results.}

\label{fig3} 

\end{figure}

If the total far-infrared-derived SFR of the clusters is normalised by total
(luminous+dark) mass, we have a simple method to compare the activity in
different environments, and the evolution of the cluster SFR budget over time
(e.g.~\citealt{Geach06}). The mass estimates for these clusters are indirectly
inferred from their optical richnesses (see Section~\ref{cluster}), which
gives the range $\sim$1.5--16$\times10^{14}$\,M$_\odot$ -- however, the
conversion between optical richness and mass is highly uncertain, and the true
masses are likely to be at the lower end of this range. Even this might be an
over-estimate of the total cluster mass. For example, to achieve a similar
surface density of clusters in the Millennium Simulation
\citep{Springel05}, 
requires us to be probing to a mass limit of $\log(M_{\rm
halo}/\mathrm{M}_\odot)\gtrsim 13.5$.

Fig.~\ref{fig3} shows the
average total SFR in our clusters compared to
other infrared-derived rates in other clusters over $0<z<0.6$,
although we note that given the large uncertainties on the H-ATLAS
cluster masses and the cool SED we have adopted, we consider
our point to be a lower limit.  We note that if other infrared studies at
similar redshifts and depths assume M82- or Arp\,220-like templates
for their cluster member galaxies when cooler H-ATLAS-type
SEDs are more appropriate, then they may well be
overestimating the level of star formation activity in those clusters.  
Although there is significant cluster-to-cluster scatter, in general
there has been strong evolution of the cluster SFR (see \citealt{Geach06}). This is
consistent with the scenario that there has been a sharper drop-off in the
star formation activity of clusters since $z\sim1$ than occurs in the field,
probably related to the build up of virialised structures hostile to on-going
activity and gas cooling over this period.

\vspace{-0.5cm}
\subsection{The nature of the optically unidentified far-infrared
  sources around the clusters}\label{unid}

Although the optically unidentified sources within $5'$ of the
clusters do not contribute significantly to the excess signal seen in
Section~\ref{sec1} and appear to lie at typically
higher redshifts ($z\sim1$) on average than the optically identified H-ATLAS
sources (Section~\ref{excess}) -- these results suggest that strong gravitational lensing by the
cluster potential is not a major contributor to the detected excess
signal. Still, they represent an interesting sample, since they
could contain strongly lensed galaxies. They thus provide potential opportunities to study the properties of
intrinsically fainter far-infrared sources at high-$z$ than would otherwise be
possible. It is possible that some of these galaxies could be galaxies at or
below the cluster redshift, but are very highly obscured, although we
note that \citet{Dunne10} do not see evidence for a significant
population of optically-faint
low-$z$ sources.  As a simple test for
this, we consider a prototypical ultraluminous infrared galaxy (ULIRG) with
extreme reddening at UV-optical wavelengths: Arp\,220
($A_V>15$\,mag, with $A_V=15$\,mag in the most central 300\,pc and
much higher in the nuclei; Vermaas \& Van der Werf in preparation). By
redshifting this template to the cluster redshifts, and normalising it to the
observed 250\,$\mu$m flux (the most sensitive band with the best angular
resolution) we can assess whether we would have detected its optical
counterpart by convolving the optical portion of the SED with, say, an
$r$-band filter. If the predicted optical flux is above the limit of the
observations of this field, then the galaxy would require even further
extinction on top of the Arp\,220 template or a different k-correction,
implying a different (higher) redshift to the cluster. We find that all of the
galaxies without current optical identifications are classified as high-$z$
sources using this method, with predicted $r_{\rm SDSS}$ at least one
magnitude brighter than the 22.4\,mag SDSS limit. The 451
optically unidentified H-ATLAS sources are thus candidate lensed sources.
Although the amplification factor falls off rapidly with radius for all
sensible mass profiles, it is worth noting that the Einstein radius for these
clusters is expected to be $\sim$20--40$''$ for a mass range of
1--$5\times10^{14}$\,M$_{\odot}$.  Therefore, only a fraction of the
451 optically unidentified H-ATLAS sources are expected to be highly
magnified. 

\begin{figure*}
\epsfig{file=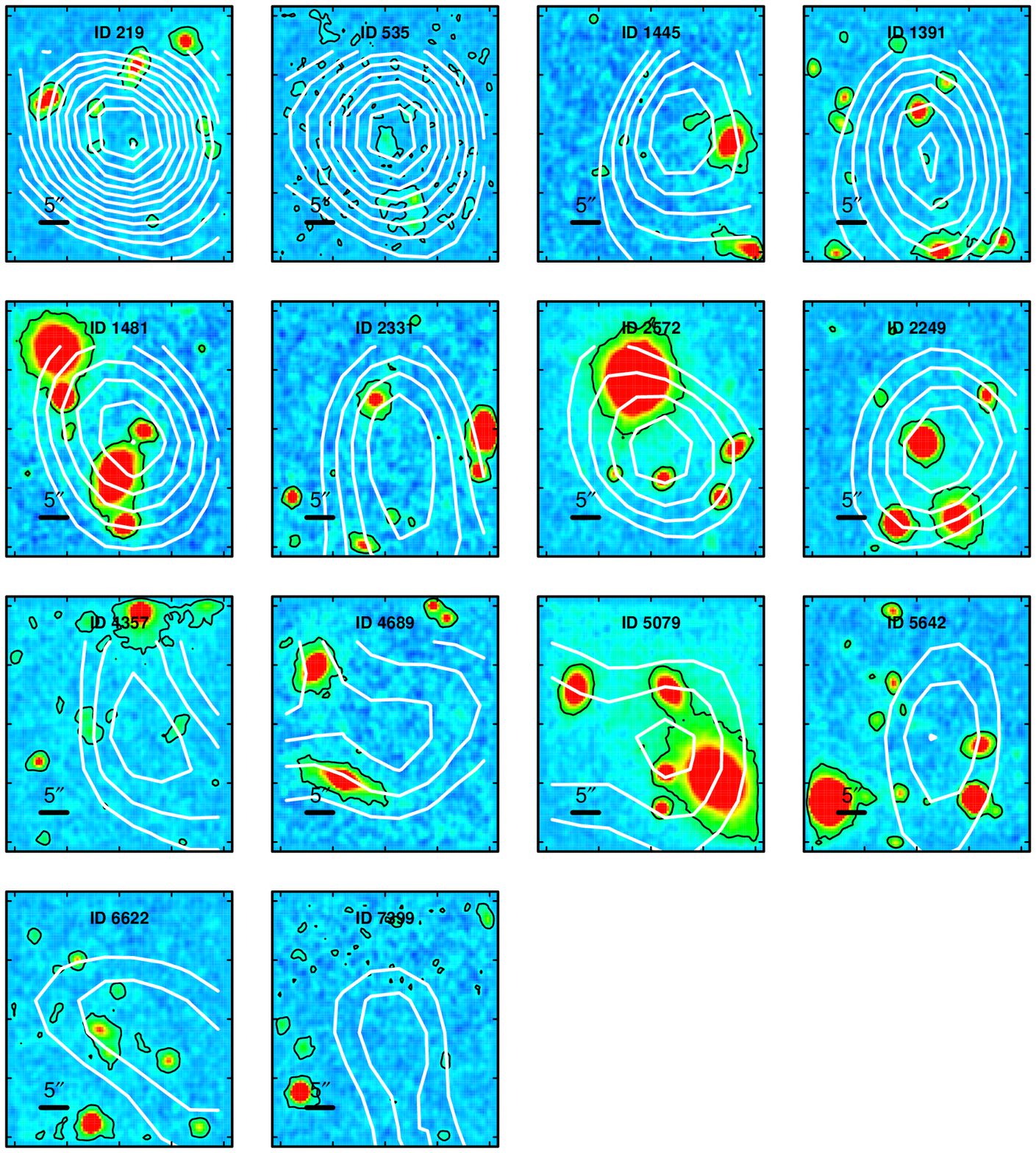,width=1\textwidth,angle=270}
\caption{Co-added $40''\times40''$ $gri$ images of the 14 strong lens
  candidate H-ATLAS sources for further study and follow-up (centred on
  the 250\,$\mu$m positions, with 250\,$\mu$m contours starting at 3\,$\sigma$
  and increasing in steps of 1\,$\sigma$) within 1$'$ ($\simeq0.2$\,Mpc) of the cluster
  centres in order of decreasing 250\,$\mu$m flux density (\textit{Left-Right,
    Top-Bottom}).}
\label{fig4}
\end{figure*}

Thus, from the 451 candidate lensed sources, we have
singled out those sources lying within 1$'$ ($\simeq0.2$\,Mpc) of the cluster centres for further
study, amounting to 14 strong lens candidates (see Table~\ref{tab}).
The mean colours of the 14 lens candidates are
  typical of the colours of the optically unidentified H-ATLAS sources
  and are consistent with being high-redshift sources: $\left< S_\mathrm{250}/S_\mathrm{350}\right> =
1.3\pm0.1$; $\left< S_\mathrm{500}/ S_\mathrm{350}\right> =
0.5\pm0.1$.
In Fig.~\ref{fig4} we present co-added {\it gri} images of the
sources, with 250$\mu$m flux density contours overlaid.

\begin{table*} 
	\caption{A list of H-ATLAS strong lens candidates
          within $1^{\prime}$ of cluster cores, ranked in order of
          decreasing 250\,$\mu$m flux density (note that these
          particular sources are all detected at $\gsim5\,\sigma$ at
          250\,$\mu$m).  The mean colours of the lens candidates are
  typical of the colours of the optically unidentified H-ATLAS sources
  and are consistent with being high-redshift sources:  $\left< S_\mathrm{250}/S_\mathrm{350}\right> =
1.3\pm0.1$; $\left<S_\mathrm{500}/S_\mathrm{350}\right> =
0.5\pm0.1$ (errors represent the error on the mean).}\label{tab}
	\begin{center}
	\begin{tabular}{llrrrcrr} 
	\hline 
IAU identifier & SDP ID & RA & Dec. & $r_{\rm c}$ & $S_{250}$ & $S_{250}/\mathrm{S}_{350}$ & $S_{500}/S_{350}$\cr 
 & & [$^{\rm h}$\,$^{\rm m}$\,$^{\rm s}$] & [$^\circ$\,$'$\,$''$] & [$'$] & [mJy] \cr
\hline
HATLAS\,J091354.6--004539 & SDP.219 & 09:13:54.7 & $-$00:45:39.6 & 0.93 & $91.8\pm6.7$ & $1.1\pm0.1$
& $0.6\pm0.1$ \\
HATLAS\,J090620.3+013112 & SDP.535 & 09:06:20.3 & 01:31:12.1 & 0.91 & $72.4\pm6.7$ & $1.2\pm0.2$ &
$0.6\pm0.2$ \\
HATLAS\,J091130.9--002227 & SDP.1445 & 09:11:30.9 & $-$00:22:27.4 & 0.68 & $59.6\pm6.9$ &
$1.2\pm0.2$ & $0.7\pm0.2$ \\
HATLAS\,J090142.6+012128 & SDP.1391 & 09:01:42.6 & 01:21:28.7 & 0.21 &$58.3\pm6.7$ & 
$1.2\pm0.3$ & $0.5\pm0.2$ \\
HATLAS\,J091231.4--000703 & SDP.1481 & 09:12:31.4 & $-$00:07:03.5 & 0.67 & $56.9\pm6.8$ &
$1.3\pm0.3$ & $0.4\pm0.2$ \\
HATLAS\,J091149.4--000424 & SDP.2331 & 09:11:49.4 & $-$00:04:25.0 & 0.67 & $49.4\pm6.8$ &
$1.8\pm0.6$ & $0.1\pm0.3$\\
HATLAS\,J091014.8--005024 & SDP.2572 & 09:10:14.9 & $-$00:50:24.4 & 0.43 & $48.2\pm6.8$ &
$1.2\pm0.3$ & $0.5\pm0.2$\\
HATLAS\,J090405.7+014443 &  SDP.2249 & 09:04:05.7 & 01:44:43.22 & 0.30
& $46.8\pm6.9$ & $0.6\pm0.1$ & $0.8\pm0.2$ \\
HATLAS\,J091233.8--004337 & SDP.4357 & 09:12:33.9 & $-$00:43:37.2 & 0.72 & $40.4\pm6.8$ &
$1.2\pm0.3$ & $0.4\pm0.3$\\
HATLAS\,J091129.1--002237 & SDP.4689 & 09:11:29.2 & $-$00:22:37.4 & 0.21 & $37.8\pm6.8$ &
$1.0\pm0.3$ & $0.7\pm0.3$\\
HATLAS\,J091309.6--003939 & SDP.5079 & 09:13:09.6 & $-$00:39:39.1 & 0.44 & $37.8\pm6.8$ &
$1.7\pm0.7$ & $0.6\pm0.5$\\
HATLAS\,J090143.0+012224 & SDP.5642 & 09:01:43.0 & 01:22:24.7 & 0.81 & $36.1\pm7.0$ & $2.6\pm1.6$ & $-0.2\pm0.7$\\
HATLAS\,J091342.4--004614 & SDP.6622 & 09:13:42.5 & $-$00:46:14.5 & 0.93 & $34.9\pm6.8$ &
$1.3\pm0.5$ & $0.6\pm0.4$ \\
HATLAS\,J090907.6--001209 & SDP.7399 & 09:09:07.6 & $-$00:12:09.9 & 0.89 & $33.4\pm6.7$ &
$0.7\pm0.2$ & $0.8\pm0.3$ \\
\cr
\hline
 \end{tabular} 
\end{center}
\end{table*}

\vspace{-0.7cm}
\section{Summary \& Conclusions}\label{summary}

We have detected a significant excess of optically identified far-infrared
sources within $\sim1.2$\,Mpc of the centres of optically-selected clusters of
galaxies with $\left<z\right>\sim 0.25$ in the SDP H-ATLAS field. Assuming
that the excess signal is completely dominated by star formation (rather than
AGN), the far-infrared excess corresponds to an average SFR of
$\approx7$\,M$_\odot$\,yr$^{-1}$ per cluster. 

The average cluster far-infrared SFR is consistent with mass-normalised SFRs from
previous work. If the observed SFR in these clusters is maintained over the
3\,Gyr since $z=0.25$, then the activity could contribute
$\sim2\times10^{10}$\,M$_{\odot}$ of new stellar mass in the clusters
-- enough to construct a typical S0 bulge. Our average
integrated SFR for the clusters can be considered a
lower-limit, since we expect additional contribution from obscured sources
below the sensitivity limit of the H-ATLAS observations. This evolution is
necessary for the observed increase in the fraction of massive (bulge
dominated) lenticular galaxies in the cores of clusters over the same time
period.

Finally, we have determined that the optically unidentified H-ATLAS sources
within 5$'$ ($\simeq1.2$\,Mpc) of the cluster cores are higher-redshift
background sources, some of which could be strongly lensed by the cluster, and
we have compiled a list of lensed candidates for further study. The future
full H-ATLAS coverage will be sufficiently large that this analysis can be
repeated and extended as a function of cluster redshift and mass. In
addition, this analysis could be applied to targetted SPIRE
observations of galaxy clusters.

\vspace{-0.7cm}
\section{Acknowledgments}\label{ack}

We thank the referee, Pierre-Alain Duc, for
his suggestions which helped to improve the paper, and also Tracy Webb
for useful discussions. KEKC and JEG acknowledge support from the the endowment of the Lorne Trottier
Chair in Astrophysics and Cosmology at McGill, the National Science and Engineering
Research Council of Canada, and the UK Science and Technology
Facilities Council (STFC).  KEKC also acknowledges the Centre of Research in Astrophysics of
Qu\'{e}bec for a fellowship.  ACE, IRS, and RJI acknowledge support
from STFC. GDZ acknowledges financial contribution from the
agreement ASI-INAF I/009/10/0.  The
\textit{Herschel}-ATLAS is a project with \textit{Herschel}, which is an ESA
space observatory with science instruments provided by European-led Principal
Investigator consortia and with important participation from NASA.

\setlength{\bibhang}{2.0em}


\begin{thebibliography}{50}
\setlength{\itemindent}{-2.5em}
\bibitem[Amblard et al.(2010)]{Amblard10}Amblard A. et al., 2010, A\&A, 518, 9L
\bibitem[Abazajian et al.(2009)]{Abazajian09}Abazajian K. N. et al., 2009, ApJS, 182, 543
\bibitem[Balogh et al.(2001)]{Balogh01}Balogh M.L., Christlein D., Zabludoff A.I., Zartsky D., 2001, ApJ, 557, 117
\bibitem[Biviano et al.(2004)]{Biviano04}Biviano A. et al., 2004,  A\&A, 425, 33
\bibitem[Braglia et al.(2011)]{Braglia10}Braglia F.G. et al., 2011,
  MNRAS, 412, 1187
\bibitem[Clowe, Gonzalez \& Markevitch(2004)]{Clowe04}Clowe D., Gonzalez A. \& Markevitch M., 2004, ApJ, 604, 596
\bibitem[Cowie et al.(2004)]{Cowie04}Cowie L.L., Barger A.J., Fomalont E.B., Capak P., 2004, ApJ, 603, L69
\bibitem[de Propris et al.(1999)]{dePropris99} de Propris R., Stanford S.A., Eisenhardt P.R., Dickinson M., Elston R., 1999, AJ, 118, 719
\bibitem[Dressler et al.(1997)]{Dressler97}Dressler A., Oemler A. Jr.,
  Butcher H.R., Gunn J.E., 1994, ApJ, 430, 107
\bibitem[Duc et al.(2000)]{Duc00}Duc P.-A., Brinks E., Springel V.,
  Pichardo B., Weilbacher P. Mirabel I.F., 2000, AJ, 120, 1238
\bibitem[Duc et al.(2002)]{Duc02}Duc P.-A., Poggianti B., Fadda D.,
  Elbaz D., Flores H., Chanial P., Franceschini A., Moorwood A.,
  Cesarsky C., 2002, A\&A, 382, 60
\bibitem[Duc et al.(2004)]{Duc04}Duc P.-A., Fadda D., Poggianti B., 
Elbaz D., Biviano A., Flores H., Moorwood A., Franceschini A.,
Cesarsky C., 2004, in IAU Colloq. 195, Outskirts of Galaxy Clusters: Intense Life
in the Suburbs, ed. A. Diaferio (San Francisco: ASP), 347
\bibitem[Dunne et al.(2011)]{Dunne10}Dunne L. et al., 2011, MNRAS
  submitted (arXiv:1012.5186)
\bibitem[Edge et al.(2010)]{Edge10}Edge A.C. et al., 2010, A\&A, 518, 47L
\bibitem[Eales et al.(2010a)]{Eales10}Eales S. et al., 2010a, PASP, 122,
  499
\bibitem[Eales et al.(2010b)]{Eales10Hermes}Eales S. et al., 2010b, A\&A,
  518, L23
\bibitem[Egami et al.(2010)]{Egami10}Egami E. et al., 2010, A\&A, 518, L12 
\bibitem[Fadda et al.(2000)]{Fadda00}Fadda D., Elbaz D., Duc P.-A.,
  Flores H., Franceschini A., Cesarsky C. J., Moorwood A.F.M., 2000,
  A\&A, 361, 827
\bibitem[Finn et al.(2010)]{Finn10}Finn R.A. et al., 2010, ApJ, 720, 87
\bibitem[Geach et al.(2006)]{Geach06}Geach J.E. et al., 2006, ApJ, 649, 661
\bibitem[Geach et al.(2009)]{Geach09}Geach J.E., Smail I., Moran S.M.,
  Treu T., Ellis R.S., 2009, ApJ, 691, 783
\bibitem[Geach et al.(2011)]{Geach11}Geach J.E., Murphy D.N. \& Bower
  R.G., 2011, MNRAS in press (arXiv: 1101.4585)
\bibitem[Griffin et al.(2010)]{Griffin10}Griffin M.J. et al. 2010, A\&A, 518, L3
\bibitem[Gunn et al.(1998)]{Gunn98}Gunn J. E. et al., 1998, AJ, 116,
  3040
\bibitem[Haines et al.(2009)]{Haines09}Haines C.P., Smith G.P., Egami
  E., Okabe N., Takada M., Ellis R.S., Moran S.M., Umetsu K., 2009, MNRAS, 396, 1297
\bibitem[Ibar et al.(2010)]{Ibar10}Ibar E. et al., 2010, MNRAS, 409, 38
\bibitem[Kennicutt(1998)]{Kennicutt98}Kennicutt R.C., 1998, ARA\&A,
  36, 189
\bibitem[Kocevski et al.(2010)]{Kocevski10}Kocevski D.D. et al., 2010,
  ApJ submitted (arXiv:1009.2750)
\bibitem[Kodama \& Bower(2003)]{Kodama03}Kodama T. \& Bower R., 2003,
  MNRAS, 346, 1
\bibitem[Kodama \& Smail(2001)]{Kodama01}Kodama T. \& Smail I., 2001,
  MNRAS, 326, 637
\bibitem[Longair \& Seldner(1979)]{Longair79}Longair M.S. \& Seldner M.,
  1979, MNRAS, 189, 433 
\bibitem[Markevitch(2006)]{Markevitch06}Markevitch M., 2006,
  Proceedings of 
the The X-ray Universe 2005 (ESA SP-604). 26-30 September 2005, El Escorial, Madrid, Spain. Editor: A. Wilson, p.723
\bibitem[Metcalfe et al.(2003)]{Metcalfe03}Metcalfe L. et al., 2003,
  A\&A, 407, 791
\bibitem[Metcalfe, Fadda \& Biviano(2005)]{Metcalfe05}Metcalfe L., Fadda D. \& Biviano A., 2005, Space Science Reviews, 119, 425
\bibitem[Meusinger et al.(2000)]{Meusinger00}Meusinger H., Brunzendorf J., Krieg R., 2000, A\&A, 363, 933
\bibitem[Murphy, Geach \& Bower(2010)]{MGB10}Murphy D., Geach J.E. \& Bower R., 2010, MNRAS submitted
\bibitem[Negrello et al.(2010)]{Negrello10}Negrello M. et al., 2010,
  Science, 330, 800
\bibitem[Neistein, van den Bosch \& Dekel(2006)]{Neistein06}Neistein
  E., van den Bosch F.C. \& Dekel A., 2006, MNRAS, 372, 933
\bibitem[Pascale et al.(2010)]{Pascale10}Pascale E., et al., 2010,
  MNRAS submitted (arXiv:1010.5782)
\bibitem[Pilbratt et al.(2010)]{Pilbratt10}Pilbratt G.L. et al., 2010,
  A\&A, 518, L1
\bibitem[Poggianti et al.(1999)]{Poggianti99}Poggianti B.M.,
  Smail I., Dressler A., Couch W.J., Barger A.J., Butcher H., Ellis
  R.S., Oemler A., 1999, ApJ, 518, 576
\bibitem[Poglitsch et al.(2010)]{Poglitsch10}Poglitsch A. et al.,
2010, A\&A, 518, 2L
\bibitem[Rawle et al.(2010)]{Rawle10}Rawle T.D. et al., 2010, A\&A,
  518, 14L
\bibitem[Rigby et al.(2011)]{Rigby10}Rigby J.R. et al.\ 2011, MNRAS
  in press (arXiv:1010.5787)
\bibitem[Smail et al.(2007)]{Smail07}Smail I. et al., 2007, ApJ, 654, 33L
\bibitem[Smith et al.(2011)]{Smith10}Smith D.J.B. et al.\ 2011,
  MNRAS in press (arXiv:1007.5260)
\bibitem[Spergel et al.(2003)]{Spergel03}Spergel D.N. et al., 2003,
  ApJS, 148, 175
\bibitem[Springel et al.(2005)]{Springel05}Springel V. et al., 2005,
  Nat., 435, 629
\bibitem[Stott et al.(2007)]{Stott07}Stott J.P., Smail I., Edge A.C.,
  Ebeling H., Smith G.P., Kneib J.-P., Pimbblet K.A., 2007, ApJ, 661,
  95
\bibitem[Temporin et al.(2009)]{Temporin09}Temporin S. et al., 2009,
  AN, 330, 915 
\bibitem[Vulcani et al.(2011)]{Vulcani10}Vulcani B. et al., 2011,
  MNRAS, 412, 246
\bibitem[Wardlow et al.(2010)]{Wardlow10}Wardlow J.L. et al., 2010,
  MNRAS, 401, 2299
\bibitem[Yee \& Ellingson(2003)]{Yee03}Yee H.K.C. \& Ellingson E.,
  2003, ApJ, 585, 215
\bibitem[York et al.(2000)]{York00}York D.G. et al.\ 2000, AJ, 120,
  1579
\end{thebibliography}
\end{document}